\documentclass{article}

\usepackage{arxiv}

\usepackage[utf8]{inputenc} 
\usepackage[T1]{fontenc}    
\usepackage{hyperref}       
\usepackage{url}            
\usepackage{booktabs}       
\usepackage{amsfonts}       
\usepackage{nicefrac}       
\usepackage{microtype}      
\usepackage{lipsum}		
\usepackage{graphicx}
\usepackage[numbers]{natbib}
\usepackage{doi}
\usepackage{pifont}
\usepackage{array}
\usepackage{pifont}
\renewcommand{\arraystretch}{1.4}
\usepackage{xcolor}
\usepackage{adjustbox}

\title{Enhancing Software Supply Chain Security through STRIDE-Based Threat Modelling of CI/CD Pipelines}

\author{Sowmiya Dhandapani \\
	Independent Cyber Security Researcher\\
	Abu Dhabi, United Arab Emirates \\
    \texttt{https://sow28mi.github.io/} \\
	\texttt{sow28mi@gmail.com} \\
}

\date{}

\renewcommand{\headeright}{}
\renewcommand{\undertitle}{}

\hypersetup{
pdftitle={A template for the arxiv style},
pdfsubject={q-bio.NC, q-bio.QM},
pdfauthor={Sowmiya Dhandapani},
pdfkeywords={Software Supply Chain Security, CI/CD Pipeline, Threat Modeling, STRIDE, DevSecOps, Security-as-Code, SLSA Framework, SSDF, Security Control Mapping},
}

\begin{document}
\maketitle
\begin{abstract}
	With the increasing adoption of Continuous Integration and Continuous Deployment (CI/CD) pipelines, securing software supply chains has become a critical challenge for modern DevOps teams. This study addresses these challenges by applying a structured threat modeling approach to identify and mitigate risks throughout the CI/CD lifecycle. By modeling a representative pipeline architecture—incorporating tools such as GitHub, Jenkins, Docker, and Kubernetes—and applying the STRIDE framework, we systematically analyze vulnerabilities at each stage, from source code management to deployment. Threats are documented and mapped to comprehensive security controls drawn from standards like NIST SP 800-218 (SSDF), OWASP Top 10 CI/CD risks, and the SLSA framework. Controls are further evaluated against SLSA maturity levels to assess improvements in trust and provenance. To operationalize these findings, the study outlines a practical security toolchain integration strategy grounded in Security-as-Code and Shift-Left/Shield-Right principles, enabling automated, enforceable security across the pipeline. This approach provides a pragmatic roadmap for enhancing CI/CD pipeline security against evolving software supply chain threats.
\end{abstract}


\section{Introduction}
The increasing reliance on Continuous Integration and Continuous Deployment (CI/CD) pipelines in modern software development has introduced new attack surfaces for adversaries to exploit. High-profile incidents such as the SolarWinds and Codecov breaches have demonstrated the catastrophic impact of supply chain attacks, where attackers compromise build and deployment mechanisms to inject malicious code or tamper with artifacts distributed downstream. These attacks bypass traditional perimeter defenses, making the software delivery pipeline itself a critical target.

Despite growing awareness, there remains a lack of structured methodologies to proactively identify and mitigate threats within CI/CD pipelines. Traditional security audits often overlook the intricate trust relationships, automated processes, and integration points that make up these pipelines. Without a comprehensive threat model, organizations struggle to implement effective controls and validate the integrity of their software supply chains.

This paper addresses this gap by applying the STRIDE threat modeling framework—originally proposed by Microsoft—to a typical CI/CD pipeline. We systematically identify threats at each stage of the pipeline, from source code management and build automation to artifact storage and deployment. For each threat identified, we propose targeted security controls designed to mitigate risk and improve the resilience of the pipeline. Additionally, we relate these controls to the Supply-chain Levels for Software Artifacts (SLSA) framework, which provides a graduated model for software supply chain security maturity.

Our contributions are threefold, which offer both a theoretical foundation and practical guidance for organizations seeking to secure their CI/CD pipelines against emerging software supply chain threats.
\begin{enumerate}
  \item We present a detailed threat model of a CI/CD pipeline using the STRIDE methodology.
  \item We identify and map security controls to each threat, categorized by STRIDE threat types.
  \item We analyze the alignment of these mitigations with the SLSA framework to demonstrate maturity and traceability.
\end{enumerate}

\section{LITERATURE REVIEW}
The security of software supply chains has garnered increased attention in recent years, particularly as adversaries shift their focus from direct application exploitation to indirect compromise of build systems and deployment pipelines. As modern software development increasingly embraces Continuous Integration and Continuous Deployment (CI/CD) pipelines, ensuring the security of the software supply chain has become a critical challenge. Historically, much of the research focused on securing individual components of the pipeline, particularly on maintaining integrity. However, with the growing sophistication of adversaries, the scope of security in CI/CD pipelines must address a wider range of threats, including those to confidentiality, availability, and overall system resilience. This section reviews existing efforts in CI/CD threat modeling, software supply chain frameworks, and SLSA adoption challenges.

\subsection{The Escalating Threat Landscape in Software Supply Chains}
The software supply chain has become an increasingly attractive target for adversaries. Attacks like SolarWinds, Codecov, and the recent XZ Utils backdoor underscore how compromising the CI/CD pipeline — rather than the application itself — can have widespread impact. These incidents have catalyzed a wave of research into software supply chain security, particularly around securing the automation systems that build, test, and deploy software.
\subsection{Existing Frameworks: Provenance and Integrity}
Several frameworks have emerged to address artifact trust and integrity. The SLSA framework~\cite{1} introduces progressive levels of assurance for build provenance, while in-toto and TUF~\cite{2} provide cryptographic guarantees for software artifacts. These efforts are essential, yet they presume a trusted CI/CD infrastructure, leaving the underlying pipeline risks underexplored.
\subsection{Threat Modeling of CI/CD Pipelines}
Reichert and Obelheiro ~\cite{3} introduced an integrity-focused threat model for software development pipelines based on the STRIDE methodology. Their approach identifies a variety of pipeline-specific threats, including unauthorized code injection, build tampering, and exposure of secrets. While their work provides a foundational model, it stops short of systematically mapping threats to actionable mitigations, which is a focus of our work.
\subsection{Software Supply Chain Security Landscape}
A systematic literature review by the same authors~\cite{4} have shown that public CI/CD workflows suffer from widespread security misconfigurations, including overprivileged tokens, leaked secrets, and unsafe third-party actions. Their findings highlight a gap in end-to-end threat modelling that accounts for modern DevOps practices.
\subsection{Adoption Challenges in the SLSA Framework}
Tamanna et al.~\cite{5} analyzed over 1,500 GitHub issues and community discussions related to the implementation of the SLSA framework. Their study highlighted key challenges such as lack of toolchain support, ambiguous specifications, and organizational inertia. These insights underscore the importance of making security controls both technically feasible and operationally realistic—an aspect our control mapping attempts to address.
\subsection{Automation of Provenance and Controls}
ARGO-SLSA, introduced by Thariq and Ekanayake~\cite{6} demonstrates how Kubernetes-native tools can be used to automatically generate SLSA-compliant provenance in CI/CD workflows. This research assumes that the pipeline is properly configured and not compromised which indicates that most existing tools and models are reactive, focusing on output verification rather than preventive, infrastructure-level threat modeling.
\subsection{Human and Organizational Challenges}
Kalu et al.~\cite{7} conducted interviews with industry practitioners and found that despite the availability of tools for signing, auditing, and validation, widespread adoption is hindered by a lack of developer awareness and cultural resistance. These insights are relevant to our work, particularly in emphasizing the need for clear, actionable controls that are easily adoptable in real-world pipelines.

Hamer et al.~\cite{8}, who conducted a broad review of the software supply chain security research landscape and emphasized the lack of systematic approaches to secure build and deployment systems. Their work outlines several future research directions—among them, the need for deeper integration of threat modeling methodologies with CI/CD infrastructure—which directly motivates the contribution of this paper.

Despite the growing body of literature and tooling, there remains a significant gap in comprehensive, end-to-end threat modelling of CI/CD pipelines. Current approaches either focus solely on artifact integrity or provide generalized DevSecOps checklists without structured methodologies. Few papers explicitly map threats across all pipeline stages — from code commit to deployment — or align them with frameworks like STRIDE to derive actionable security controls.

This paper addresses that gap by:
\begin{enumerate}
    \item Applying STRIDE-based threat modeling across each CI/CD stage — covering the full threat landscape beyond integrity.
    \item Identifying specific threats and their mitigations in a proactive and structured manner.
    \item Mapping security controls to SLSA levels, showing how threat modelling can operationalize and strengthen existing supply chain frameworks.
\end{enumerate}

\section{METHODOLOGY}
This study adopts a structured threat modeling approach to identify and mitigate software supply chain risks within Continuous Integration/Continuous Deployment (CI/CD) pipelines. The methodology consists of four key phases: pipeline scoping, threat modelling using STRIDE, security control mapping, and validation through comparative analysis.

\subsection{CI/CD Pipeline Scoping}
We begin by defining a representative CI/CD pipeline architecture that reflects common industry practices. The pipeline includes stages such as source code management, build, test, artifact storage, image signing, containerization, and deployment. Tools such as GitHub, Jenkins, Docker, and Kubernetes are considered in the model to reflect real-world implementations. Each pipeline stage is treated as a potential attack surface.
\subsection{Threat Modeling Using STRIDE}
To systematically analyze threats across the pipeline, we apply the STRIDE threat modeling methodology—focusing on Spoofing, Tampering, Repudiation, Information Disclosure, Denial of Service, and Elevation of Privilege. Each stage of the CI/CD pipeline is assessed for vulnerabilities associated with these threat categories. Threats are documented using a Data Flow Diagram (DFD) to visualize assets, trust boundaries, and data exchanges.
\subsection{Security Control Identification}
For each threat identified through STRIDE, corresponding mitigating controls are proposed. Controls are mapped using a combination of:
\begin{itemize}
    \item NIST SP 800-218 (SSDF) Security Control references
    \item OWASP Top 10 CI/CD Risks
    \item Supply-Chain Level for Software Artifacts (SLSA) framework recommendations
    \item Industry best practices for securing DevOps workflows
\end{itemize}
Each control is aligned to the affected pipeline component and categorized by control type (preventive, detective, corrective).
\subsection{Alignment with SLSA Framework}
To assess the maturity of the proposed security controls, each identified control is mapped against the four levels of the SLSA framework. This helps evaluate how the controls improve trust, traceability, and provenance within the pipeline.
\subsection{Security Toolchain Integration for Pipeline Hardening}
To operationalize the proposed threat model and security controls, a practical integration strategy is outlined with relevant tools and configurations into CI/CD pipelines. The objective is to translate the identified risks and corresponding controls into enforceable security mechanisms using widely adopted DevSecOps tooling.
This integration approach is guided by two foundational principles:
\begin{enumerate}
    \item Security-as-Code: Security checks are implemented as codified configurations or policies within the pipeline. This ensures reproducibility, versioning, and seamless enforcement alongside code changes.
    \item Shift-Left and Shield-Right: Controls are embedded early in the pipeline to prevent issues at source (shift-left) while also ensuring runtime enforcement and monitoring at later stages (shield-right).
\end{enumerate}
For each threat-to-control mapping established earlier, this phase identifies compatible tools or frameworks that enable the enforcement or verification of controls. These tools are selected based on criteria such as industry relevance, compatibility with cloud-native CI/CD environments, support for automation, and adherence to open standards.

\section{BACKGROUND}
This section introduces the foundational concepts and frameworks referenced in this study, including CI/CD pipelines, the STRIDE threat modelling methodology, and security frameworks such as SLSA and SSDF. These elements collectively form the basis for identifying, analyzing, and mitigating risks in software supply chain workflows.
\subsection{Continuos Integration and Continuous Deployment}
Continuous Integration and Continuous Deployment (CI/CD) are cornerstone practices in modern DevOps that aim to streamline software delivery by automating the build, test, and release processes. These pipelines enable rapid and reliable deployment of software, minimizing human error and increasing developer productivity. However, their interconnected and automated nature also introduces numerous security risks.

\begin{figure}[htbp]
  \centering
  \fcolorbox{black}{white}{%
    \begin{adjustbox}{left, padding=5pt}
      \includegraphics[width=1.2\textwidth,keepaspectratio]{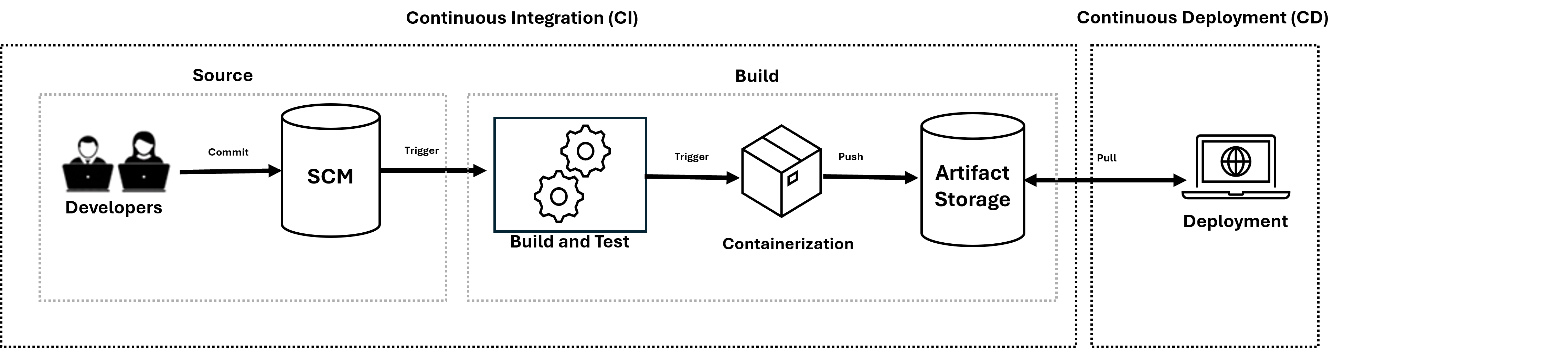}
    \end{adjustbox}
  }
  \caption{Typical CI/CD Pipeline}
  \label{fig:fig1}
\end{figure}

A typical CI/CD pipeline in Figure \ref{fig:fig1} involves the following critical stages:
\begin{itemize}
    \item \textbf{Source Code Management (SCM)}: Developers collaborate using distributed version control systems such as Git, with platforms like GitHub, GitLab, or Bitbucket facilitating code management. This stage often involves webhooks and third-party integrations, forming the initial attack surface where source code integrity can be compromised via malicious commits, unauthorized access, or poisoned dependencies.
    \item \textbf{Build and Test Automation}: Once code is committed, automated build systems (e.g., Jenkins, GitHub Actions, GitLab CI) compile the application and run test suites. This stage can include code quality analysis, unit testing, and integration testing. Compromises in build agents, insecure build scripts, or insufficient isolation can lead to tampering or injection attacks.
    \item \textbf{Containerization}:Applications are packaged into containers using tools like Docker to enable consistent deployment across environments. While containerization is a common practice, image signing is not typically included by default in CI/CD pipelines. When implemented, image signing tools such as Sigstore Cosign or Docker Notary are used to generate cryptographic signatures for container images, providing assurances about the image’s integrity and origin. These signatures help prevent unauthorized modifications and support verification during later stages of the deployment. However, many pipelines skip this step due to lack of awareness, tooling complexity, or performance trade-offs. This introduces risks such as insecure image builds, propagation of unsigned or tampered containers, and the use of unverified base images—leaving the software supply chain vulnerable to attacks.
    \item \textbf{Artifact Storage:} The compiled binaries, packages, and Docker images are stored in artifact repositories such as JFrog Artifactory, Nexus, or container registries like Docker Hub or Amazon ECR. If these storage systems are not securely configured, attackers may replace or inject malicious artifacts, leading to downstream compromise during deployment.
    \item \textbf{Deployment and Orchestration:} The final step involves deploying containers or artifacts to production environments using orchestration tools such as Kubernetes, ArgoCD, or Helm. Security misconfigurations, exposed APIs, or insufficient Role-Based Access Control (RBAC) can allow attackers to escalate privileges or disrupt services.
\end{itemize}
Each of these stages introduces unique security concerns due to the diverse tool chain integrations, complex dependency trees, and trust boundaries between components. Threat actors increasingly target CI/CD pipelines to inject malicious code, steal credentials, or exfiltrate sensitive artifacts. Recent supply chain attacks such as those on SolarWinds and Codecov exemplify how compromising a single component in the pipeline can lead to widespread breaches.
    
This underscores the importance of modeling threats specific to CI/CD workflows and enforcing robust, end-to-end security controls across the pipeline.

\subsection{STRIDE Threat Modelling}
STRIDE is a mnemonic-based threat modeling framework developed by Microsoft to help systematically identify security threats in software systems. It categorizes potential threats into six classes—Spoofing, Tampering, Repudiation, Information Disclosure, Denial of Service, and Elevation of Privilege—each representing a specific type of adversarial behavior that can compromise the confidentiality, integrity, or availability of software systems.

In the context of CI/CD pipelines, applying STRIDE allows for a granular threat analysis across each pipeline component and interaction. Each STRIDE category aligns naturally with risks introduced by automation, integration points, and complex tool chains inherent in modern DevOps environments:
\begin{enumerate}
    \item \textbf{Spoofing}: Unauthorized entities (e.g., attackers or compromised insiders) may impersonate developers, build agents, or services by exploiting weak authentication mechanisms. For example, attackers could gain access to a source code management or version control system using stolen credentials or improperly configured access tokens.
    \item \textbf{Tampering}: Threat actors may alter source code, build scripts, container images, or deployment configurations at any pipeline stage. Code repositories and artifact registries are especially vulnerable if integrity checks and signing mechanisms are absent.
    \item \textbf{Repudiation}: Without robust logging and non-repudiation controls, malicious actions (e.g., unauthorized configuration changes) may go undetected or be wrongly attributed. This impairs incident response and forensics, creating blind spots in the CI/CD audit trail.
    \item \textbf{Information Disclosure}: Sensitive data such as access keys, API secrets, or environment variables may be exposed through misconfigured repositories, insecure pipeline logs, or accidental inclusion in artifacts. Leakage of such information can be a precursor to broader supply chain attacks.
    \item \textbf{Denial of Service (DoS)}: Attackers may disrupt the CI/CD pipeline by overwhelming build servers, exhausting compute resources, or exploiting flaws in third-party plugins. This impacts delivery timelines and can halt critical security updates.
    \item \textbf{Elevation of Privilege}: Exploiting configuration weaknesses, such as overly permissive IAM roles or unprotected build agents, allows attackers to gain administrative control within the pipeline and execute arbitrary code or manipulate deployments.
\end{enumerate}
By mapping STRIDE threats to each CI/CD stage—from source code management to deployment, security architects can identify systemic weaknesses and implement targeted controls such as zero-trust principles, immutable infrastructure, cryptographic verification, and centralized audit logging. This structured analysis is crucial for preempting sophisticated software supply chain attacks and aligning CI/CD security practices with frameworks like SLSA and SSDF.

To illustrate the practical relevance of this approach, Table  \ref{tab:stride-incidents} presents a mapping of STRIDE threat categories to real-world software supply chain incidents. 

\begin{table}[ht]
\centering
\caption{Mapping STRIDE Threats to Real-World Incidents}
\label{tab:stride-incidents}
\renewcommand{\arraystretch}{1.3}
\begin{tabular}{|l|p{4.2cm}|p{7.8cm}|}
\hline
\textbf{STRIDE} & \textbf{Real-World Incident} & \textbf{Description / Reference} \\
\hline
Spoofing & Codecov Bash Uploader Attack (2021) & Malicious actor modified trusted Bash Uploader script, exfiltrating secrets from CI environments.~\cite{codecove2021}. \\
\cline{2-3}
 & CircleCI Incident (2023) & Compromised developer credentials enabled access to sensitive CI environment variables and tokens~\cite{circleci2023}. \\
\cline{2-3}
 & XcodeGhost (2015) & Developers unknowingly used a trojanized Xcode version, leading to spoofed app signing~\cite{xcodeghost2015}. \\
\hline
Tampering & SolarWinds Orion Backdoor (2020) & Build system compromise led to injection of malicious code into Orion software updates~\cite{sunburst2020}. \\
\cline{2-3}
 & PHP Git Server Breach (2021) & Attacker pushed malicious commits directly to PHP’s source repo~\cite{php2021}. \\
\cline{2-3}
 & Homebrew GitHub Attack (2021) & Attacker attempted to tamper with popular package management tools via GitHub pull requests~\cite{homebrew2021}. \\
\hline
Repudiation & Event-Stream npm Attack (2018) & Malicious code was added to a dependency without proper review; lack of logging delayed detection~\cite{eventstream2018}. \\
\cline{2-3}
 & JetBrains TeamCity Exploitation (2020) & Attribution was difficult due to insufficient audit logs and observability in CI systems~\cite{teamcity2020}. \\
\hline
Information Disclosure & Uber AWS Credentials Leak (2016) & Hardcoded AWS keys pushed to GitHub, leading to data breach of 57 million users~\cite{uber2016}. \\
\cline{2-3}
 & Heroku \& Travis CI OAuth Leak (2022) & Exposed GitHub tokens gave attackers access to CI-connected repositories~\cite{heroku2022}. \\
\cline{2-3}
 & Slack GitHub Token Exposure (2015) & GitHub integration tokens were accidentally leaked, potentially granting repo access~\cite{slack2015}. \\
\hline
Denial of Service & GitHub Actions Crypto Mining (2021) & Attackers used public repos to run unauthorized mining jobs, exhausting CI resources~\cite{githubmining2021}. \\
\cline{2-3}
 & npm Registry Incident (2022) & Malicious behavior in a core dependency caused build failures and pipeline outages globally~\cite{npm2022}. \\
\cline{2-3}
 & PyPI Flooding Attack (2022) & Malicious users spammed the registry, leading to temporary shutdowns~\cite{pypi2022}. \\
\hline
Elevation of Privilege & GitHub Actions RCE (2021) &Improper use of privileged workflows allowed untrusted PRs to execute code~\cite{githubrce2021}. \\
\cline{2-3}
 & Travis CI Misconfig (2021) & Leaked environment variables from builds allowed unauthorized privilege escalation~\cite{travisci2021}. \\
\cline{2-3}
 & Azure DevOps RBAC Misconfig (2022) & Inadequate repo/project-level RBAC led to unauthorized access and potential privilege elevation~\cite{azuredevops2022}. \\
\hline
\end{tabular}
\end{table}

These cases highlight how specific threat vectors have been exploited in the wild, reinforcing the need for proactive threat modelling across the CI/CD pipeline.

\subsection{Supply Chain Levels for Software Artifacts (SLSA)}
The Supply Chain Levels for Software Artifacts (SLSA) is a security framework initiated by Google and currently maintained by the Open Source Security Foundation (OpenSSF). It offers a structured, incremental approach for safeguarding software build and delivery processes, with the goal of preventing tampering, ensuring integrity, and improving the auditability of software artifacts.

SLSA defines four progressive levels of assurance, each building on the previous one:
\begin{enumerate}
    \item \textbf{Level 1 – Build Script Available}: The build process is scripted and documented, ensuring basic repeatability. While this offers minimal protection, it provides a starting point for improving transparency and traceability.
    \item \textbf{Level 2 – Hosted Build and Provenance}: Builds are executed on a hosted platform, and provenance metadata is generated. This metadata includes the identity of the build system and links to the source and inputs used, enabling basic tamper detection.
    \item \textbf{Level 3 – Trustworthy Builds}: The build system is hardened against attacks and is capable of producing verifiable, non-falsifiable provenance. This helps ensure that the build was not influenced or altered by a malicious actor.
    \item \textbf{Level 4 – Hermetic and Reproducible Builds}: The build process is fully hermetic (i.e., isolated from external influences) and reproducible. Independent parties can recreate the same artifact from the same source, enabling maximal trust in the supply chain.
\end{enumerate}
SLSA serves as a maturity model and implementation guide, helping organizations transition from informal, ad hoc build processes to secure, verified, and auditable pipelines.

Recent high-profile incidents such as the SolarWinds SUNBURST compromise and the UAParser.js npm hijack have accelerated the industry adoption of frameworks like SLSA, underlining the need for end-to-end artifact integrity and verifiability in software supply chains. Moreover, the framework integrates well with DevSecOps workflows and complements other standards such as NIST's SSDF.

\subsection{Secure Software Development Framework (SSDF)}
The Secure Software Development Framework (SSDF), outlined in NIST Special Publication 800-218, is a foundational set of guidelines for integrating security throughout the Software Development Life Cycle (SDLC). Developed by the U.S. National Institute of Standards and Technology (NIST), SSDF is designed to help organizations reduce the risk of software vulnerabilities and enhance the trustworthiness of software products and services.

The SSDF is structured around four primary groups of practices:
\begin{enumerate}
    \item \textbf{Prepare the Organization (PO)}: Establish and maintain secure development practices and supporting infrastructure. This includes defining security roles, providing adequate training, and implementing secure tooling for code analysis and version control.
    \item \textbf{Protect the Software (PS)}: Implement mechanisms to protect code from unauthorized access or modification. This encompasses secure coding practices, managing third-party components, and ensuring the integrity of development and build environments.
    \item \textbf{Produce Well-Secured Software (PW)}: Integrate automated testing, code reviews, static and dynamic analysis, and other quality assurance practices to identify and remediate security weaknesses during development.
    \item \textbf{Respond to Vulnerabilities (RV)}: Define procedures for receiving, reporting, analyzing, and remediating vulnerabilities in released software. This ensures prompt and effective handling of security issues throughout the software lifecycle.
\end{enumerate}
In the context of CI/CD pipelines and software supply chains, SSDF helps enforce discipline across multiple layers—ranging from secure source control to hardened build and deployment environments. 

While the SSDF offers comprehensive guidance on embedding security practices across the software development lifecycle, its implementation becomes even more effective when aligned with complementary frameworks. STRIDE, as a threat modeling methodology, enables security teams to systematically identify and assess potential threats across CI/CD workflows—from spoofing and tampering to elevation of privilege. In parallel, SLSA (Supply-chain Levels for Software Artifacts) introduces a prescriptive maturity model focused on ensuring the provenance, integrity, and verifiability of software artifacts throughout the supply chain.

Together, these frameworks provide layered and mutually reinforcing security coverage. STRIDE helps uncover systemic vulnerabilities, SLSA builds trust through traceable software builds, and SSDF grounds both by offering detailed operational practices to implement technical and procedural mitigations. While SLSA emphasizes artifact-level verifiability, SSDF addresses broader organizational aspects, such as secure culture, workforce readiness, and incident response planning. Collectively, STRIDE, SLSA, and SSDF form a multidimensional foundation for proactively preempting, detecting, and mitigating software supply chain threats in modern CI/CD environments.

\section{ANALYSIS}
This section critically analyzes the alignment and limitations of existing frameworks, particularly SLSA, in the context of securing CI/CD pipelines. It highlights the necessity of applying a structured threat modeling approach (via STRIDE+SLSA), supported by SSDF, to address software supply chain threats comprehensively.

\subsection{SLSA–STRIDE Gap Analysis}
While the Supply Chain Levels for Software Artifacts (SLSA) framework significantly advances the integrity and transparency of software build and delivery processes, it is not a comprehensive threat model. Specifically, SLSA is primarily concerned with tamper resistance, artifact provenance, and build reproducibility—key components for ensuring trust in the software supply chain. However, modern CI/CD environments are complex ecosystems involving multiple tools, identities, and external dependencies, which expose them to a broader range of threats classified under the STRIDE model. This subsection presents a comparative analysis of SLSA and STRIDE, demonstrating how their combined implementation enables a defense-in-depth approach.
\subsubsection{Limitations of SLSA When Used in Isolation}
When SLSA alone is implemented, the threat classified under the STRIDE are not completely mitigated as shown in Table \ref{tab:slsa-gap-analysis}.

\begin{table}[ht]
\centering
\caption{Evaluation of SLSA Coverage and Remaining Gaps Across Key Security Aspects}
\label{tab:slsa-gap-analysis}
\renewcommand{\arraystretch}{1.3}
\begin{tabular}{|p{4cm}|p{5.5cm}|p{5.5cm}|}
\hline
\textbf{Aspect} & \textbf{SLSA Coverage} & \textbf{Gap} \\
\hline

Identity Authentication & Assumes trust in build service identity & Does not address Spoofing across developer identities or API tokens \\
\hline

Tamper Resistance & Focuses on artifact tampering & Limited to build system – less emphasis on config tampering or SCM \\
\hline

Auditing \& Accountability & Provides provenance & Not sufficient for detailed forensic Repudiation events \\
\hline

Information Protection & Metadata integrity ensured & Secrets and credentials protection not explicitly enforced \\
\hline

Availability \& Resilience & Not a core focus & Denial of Service (DoS) risks in pipeline stages are unaddressed \\
\hline

Privilege Management & Recommends two-person review (Level 4) & Broader Elevation of Privilege risks in CI/CD roles are overlooked \\
\hline

\end{tabular}
\end{table}

\subsubsection{Mapping of  SLSA with STRIDE}

To evaluate how effectively the SLSA framework mitigates threats categorized under STRIDE, we analyze its maturity levels (L1–L4) against each threat type across CI/CD pipeline. Table~\ref{tab:slsa-stride-gap} illustrates this threat-centric mapping by indicating which STRIDE categories are fully addressed, partially mitigated, or not covered at all across the different SLSA levels. The analysis highlights the extent to which SLSA provides defensive coverage and where notable security gaps remain.

While the mapping reveals strong alignment in areas like tampering and repudiation—especially at higher levels—critical gaps persist for other threat categories. For example:
\begin{itemize}
    \item \textbf{Denial of Service (DoS)}: Attacks targeting CI runners or artifact registries are not explicitly addressed by SLSA.
    \item \textbf{Spoofing}: Identity verification is scoped only to build infrastructure, not to individual developers or API tokens.
    \item \textbf{Elevation of Privilege}: SLSA partially mitigates build-time privilege escalation but lacks broader coverage of misconfigurations or post-build lateral movement.
\end{itemize}

\begin{table}[ht]
\centering
\caption{SLSA Threat Mitigation Coverage by STRIDE Category}
\label{tab:slsa-stride-gap}
\renewcommand{\arraystretch}{1.3}
\begin{tabular}{|p{3.3cm}|p{1.5cm}|p{1.5cm}|p{1.5cm}|p{1.8cm}|p{6cm}|}
\hline
\textbf{STRIDE Threat Category} & \textbf{SLSA L1} & \textbf{SLSA L2} & \textbf{SLSA L3} & \textbf{SLSA L4} & \textbf{Gap Identified} \\
\hline

Spoofing & \ding{55} & \ding{55} & \ding{51} & \ding{51} (Partial) & Identity verification limited to build infrastructure, not developer actors \\
\hline

Tampering & \ding{51} (Partial) & \ding{51} & \ding{51} & \ding{51} & Source-side tampering still possible (pre-provenance) \\
\hline

Repudiation & \ding{51} (Partial) & \ding{51} & \ding{51} & \ding{51} & Build provenance helps, but source commit repudiation remains \\
\hline

Information Disclosure & \ding{55} & \ding{51} (Partial) & \ding{51} & \ding{51} & No controls for secrets in logs, tokens in environments \\
\hline

Denial of Service & \ding{55} & \ding{55} & \ding{55} & \ding{55} & Resource abuse and CI flooding not addressed \\
\hline

Elevation of Privilege & \ding{55} & \ding{55} & \ding{55} & \ding{51} (Partial) & Limited to hardened builders; RBAC and misconfigurations ignored \\
\hline

\end{tabular}
\end{table}

The “Gap Identified” column emphasizes areas where SLSA’s security assurances are either limited in scope or not explicitly defined.

\subsubsection{The Power of Combining SLSA with STRIDE}
This analysis reinforces the importance of augmenting SLSA with STRIDE-based threat modeling to achieve robust CI/CD pipeline security. While SLSA provides a prescriptive maturity model focused on provenance and artifact integrity, STRIDE complements it by offering a structured approach to identifying and analyzing threats across the software development lifecycle. Together, they enable:

\begin{enumerate}
    \item End-to-end threat coverage across the CI/CD lifecycle, from source to deployment.
    \item Clear identification of trust boundaries and potential attack surfaces.
    \item Mapping of SLSA’s technical controls (e.g., provenance, hermetic builds) to STRIDE categories, particularly Tampering and Repudiation.
    \item Design and implementation of layered security controls based on actual threat models and risk exposure.
\end{enumerate}

By combining SLSA’s implementation guidance with the diagnostic rigor of STRIDE, organizations can build secure-by-design, threat-resilient CI/CD systems capable of withstanding modern software supply chain attacks.

\paragraph{Example - Combined Mitigation Scenario}:
A malicious actor attempts to inject a backdoor into the codebase via a compromised developer account.

\begin{itemize}
    \item \textbf{STRIDE} flags Spoofing and Tampering risks during the threat modeling of source control systems.
    \item \textbf{SLSA Level 3 or above} ensures the final build artifact is verifiable and was generated from trusted, auditable source inputs.
    \item \textbf{Additional STRIDE-informed controls}, such as RBAC, versioned logging, and anomaly detection, mitigate Elevation of Privilege and enable post-incident traceability.
\end{itemize}

\subsection{Control Surface vs. Threat Surface Misalignment}
Another critical issue is the mismatch between where SLSA applies controls (e.g., build steps, provenance signing) and where real-world threats occur. Many high-impact supply chain attacks (e.g., SolarWinds) exploited weaknesses before build, such as compromised developer credentials, malicious commits, or dependency poisoning.
    
\begin{table}[ht]
\centering
\caption{Applicability of SLSA Controls and STRIDE Modeling Across CI/CD Stages}
\label{tab:stage-slsa-stride}
\renewcommand{\arraystretch}{1.3}
\begin{tabular}{|p{3.2cm}|p{4.5cm}|p{3.2cm}|p{3.2cm}|}
\hline
\textbf{CI/CD Stage} & \textbf{Typical Threats} & \textbf{SLSA Controls Apply?} & \textbf{Requires STRIDE Threat Modeling?} \\
\hline

Source Code & Malicious commits, secrets in code & Partial (L4) & \ding{51} \\
\hline

CI Configuration & Insecure runners, exposed tokens & \ding{55} & \ding{51} \\
\hline

Build \& Artifact Creation & Tampering, unverified tools & \ding{51} (L2–L4) & \ding{51} \\
\hline

Image Deployment & Privilege escalation, misconfigured RBAC & \ding{55} & \ding{51} \\
\hline

\end{tabular}
\end{table}

Table~\ref{tab:stage-slsa-stride} provides a stage-wise evaluation of typical threats encountered across the CI/CD pipeline and examines the extent to which SLSA controls apply at each stage. It also highlights whether additional STRIDE-based threat modeling is necessary to achieve comprehensive security. 

The analysis reveals that while SLSA offers meaningful controls in the build and artifact creation phases (especially at Levels 2–4), it provides limited or no coverage in earlier (source code) and later (deployment) stages. For example, source code manipulation and misconfigured RBAC in deployments are outside SLSA’s direct scope but pose critical risks. In contrast, STRIDE-based threat modeling remains relevant across all stages—filling the visibility and control gaps not addressed by SLSA.

This complementary view underscores the importance of combining SLSA’s maturity model with STRIDE's systematic threat identification to ensure full-spectrum CI/CD pipeline security.

\subsection{SSDF’s Role in Bridging Development Risks}
To address pre-build and developer-originated risks, such as insecure code, poor configuration hygiene, and unmanaged dependencies, we incorporate the NIST Secure Software Development Framework (SSDF), defined in SP 800-218. SSDF introduces operational practices that reinforce the left side of the CI/CD pipeline and extend security coverage upstream of the build process. Specifically, SSDF promotes:
\begin{itemize}
    \item Secure coding practices
    \item Dependency and secret scanning
    \item Secure configuration and access control policies
    \item Vulnerability disclosure and remediation workflows
\end{itemize}

When aligned with STRIDE for threat modeling and SLSA for supply chain assurance, SSDF effectively bridges the security gaps on the development side, making the entire CI / CD pipeline defensible from end to end.

\subsection{Justification for Threat Modeling Prior to Control Mapping}
Threat modeling serves as a foundational step in designing secure CI/CD pipelines. Before selecting and applying controls from frameworks like SLSA and SSDF, it is critical to first identify:
\begin{itemize}
    \item Key assets and trust boundaries
    \item Attack surfaces across pipeline stages
    \item Misconfigurations and architectural weaknesses that amplify risk
\end{itemize}
Without this preliminary analysis, organizations risk applying security controls in a reactive or checklist-driven manner—potentially overlooking critical threats while over-securing less relevant areas.

By conducting STRIDE-based threat modeling per pipeline stage first, and only then mapping mitigations from SLSA and SSDF, organizations can ensure that:
\begin{itemize}
    \item Controls are threat-driven, not checklist-driven
    \item Security investments are risk-prioritized and resource-aligned
    \item Residual risks are clearly identified, documented, and consciously accepted or mitigated
\end{itemize}

\subsection{Summary of Analysis}
This analysis highlights the limitations of relying solely on the SLSA framework and emphasizes the need for a comprehensive, layered approach to software supply chain security. Key observations include:
\begin{enumerate}
    \item \textbf{SLSA is control-focused but not inherently threat-aware:} It prescribes technical safeguards but does not explicitly model real-world attacker behavior.
    
    \item \textbf{STRIDE identifies contextual, stage-specific threats:} It surfaces risks—such as spoofing and privilege escalation—that are not fully addressed by SLSA.
    
    \item \textbf{SSDF strengthens secure development posture:} It introduces organizational and procedural safeguards that extend security coverage beyond the build system.
\end{enumerate}
In conclusion, combining SLSA, STRIDE, and SSDF creates a robust, threat-aware strategy that enhances CI/CD pipeline resilience. This integrated approach ensures complete coverage—from secure code to build provenance—while adapting to evolving software supply chain threats.

\section{PROPOSED FRAMEWORK}
Based on the preceding analysis, we propose a threat-driven CI/CD security framework that integrates STRIDE-based threat modeling, SLSA control maturity, and SSDF secure development practices across all pipeline stages. This layered defense-in-depth strategy ensures that security controls are not only present, but also contextually relevant to the threats they aim to mitigate.

The framework emphasizes mapping real-world threat scenarios to specific pipeline stages and applying appropriate controls drawn from SLSA and SSDF. By aligning security measures with identified risks, the framework promotes both efficiency and effectiveness in mitigating software supply chain threats.

\begin{figure}[htbp]
  \centering
  \includegraphics[width=\textwidth,keepaspectratio]{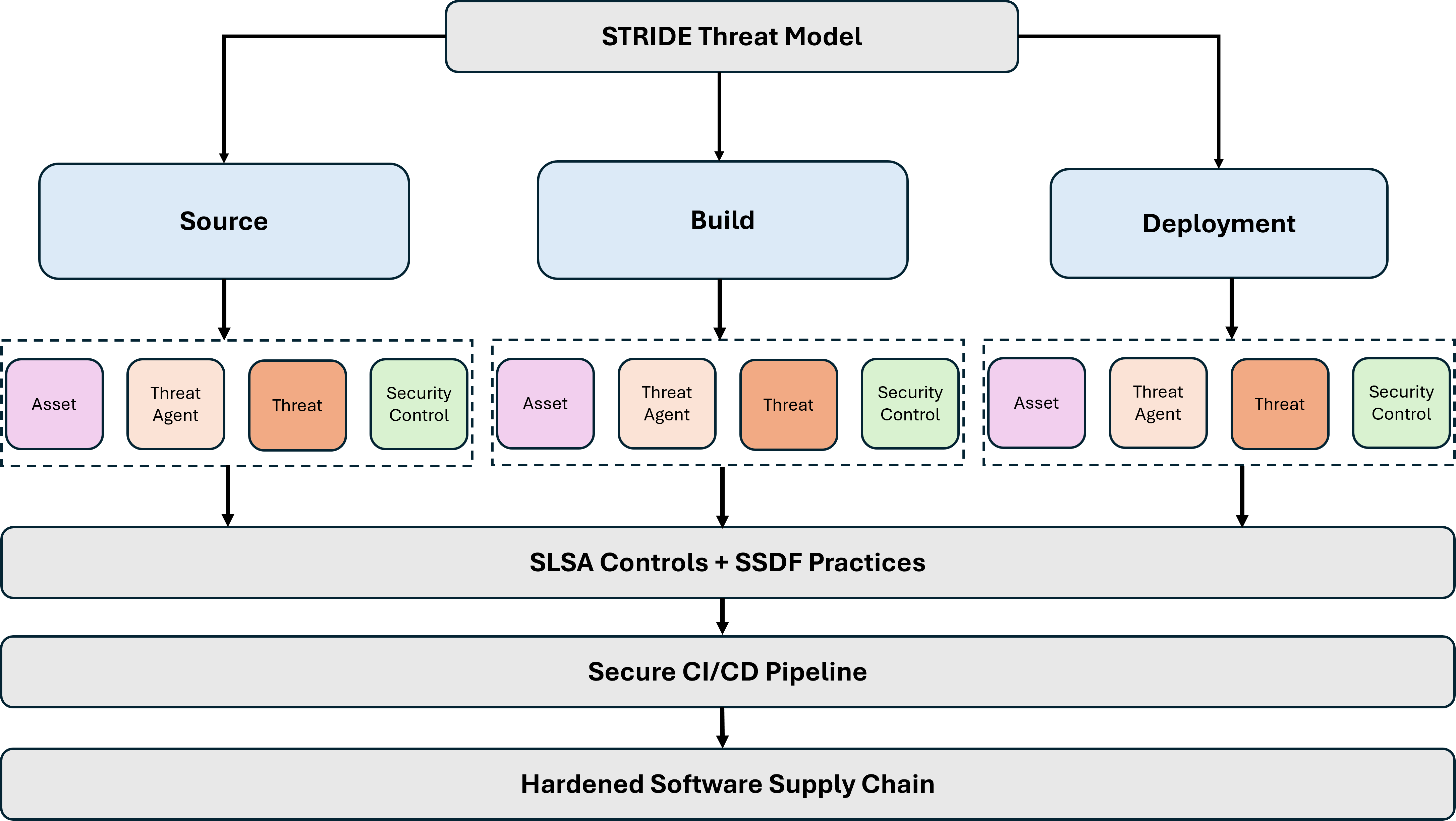}
  \caption{Proposed Composite (STRIDE+SLSA+SSDF) Framework}
  \label{fig:fig2}
\end{figure}

Figure~\ref{fig:fig2} illustrates this layered approach, showing how threats are systematically identified and addressed throughout the CI/CD pipeline—from code to build, and into deployment.

\subsection{STRIDE Threat Modelling}
The proposed framework begins by applying STRIDE-based threat modeling across the entire CI/CD pipeline—from source control and CI configuration to build and deployment stages. This enables the systematic identification of relevant threat categories—such as Spoofing, Tampering, Repudiation, and Elevation of Privilege—tailored to each stage’s assets and interactions.
\subsection{Asset–Threat Agent–Threat–Control Mapping}
For each asset involved in the CI/CD process (e.g., build servers, pipeline configuration files), potential threat agents (e.g., insider actors, compromised dependencies) are identified, along with the corresponding threats they may introduce. Security controls are then mapped to mitigate these threats, forming a traceable chain from asset to threat agent to control.
\subsection{SLSA Controls and SSDF Practices}
The threats identified through STRIDE modeling are addressed using a combination of controls from the SLSA framework and secure development practices from the SSDF:
\begin{itemize}
    \item \textbf{SLSA controls:} Verified build provenance, isolated and hermetic builds, and cryptographic attestation mechanisms.
    \item \textbf{SSDF practices:} Secure configuration (PS), code integrity verification (PW), and access control enforcement (PO).
\end{itemize}
\subsection{Secure CI/CD Pipeline Outcome}
This structured combination of threat modeling and layered controls results in a resilient CI/CD pipeline. It is equipped to withstand common attack vectors such as source code tampering, unauthorized access to build environments, and privilege escalation.
\subsection{Hardened Software Supply Chain Security}
Ultimately, this integrated approach supports the development of an end-to-end hardened software supply chain. It addresses both technical and procedural gaps surfaced through STRIDE analysis and operationalizes mitigation using SLSA and SSDF controls—achieving both visibility and verifiability throughout the pipeline.

\section{THREAT TRACEABILITY AND SECURITY MAPPING FOR THE CI/CD PIPELINE}
This section presents a detailed threat analysis derived from a Data Flow Diagram (DFD) of the end-to-end CI/CD pipeline. The objective is to identify critical assets, potential threat agents, and associated STRIDE-based threats, and to map these threats to appropriate SLSA controls and SSDF practices. This process establishes full traceability and ensures that each threat is addressed through contextual, layered security mechanisms.
\subsection{CI/CD Pipeline DFD Overview}
\begin{figure}[htbp]
  \centering
  \fcolorbox{black}{white}{%
    \begin{adjustbox}{minipage=\textwidth, center, padding=5pt}
      \includegraphics[width=\textwidth,keepaspectratio]{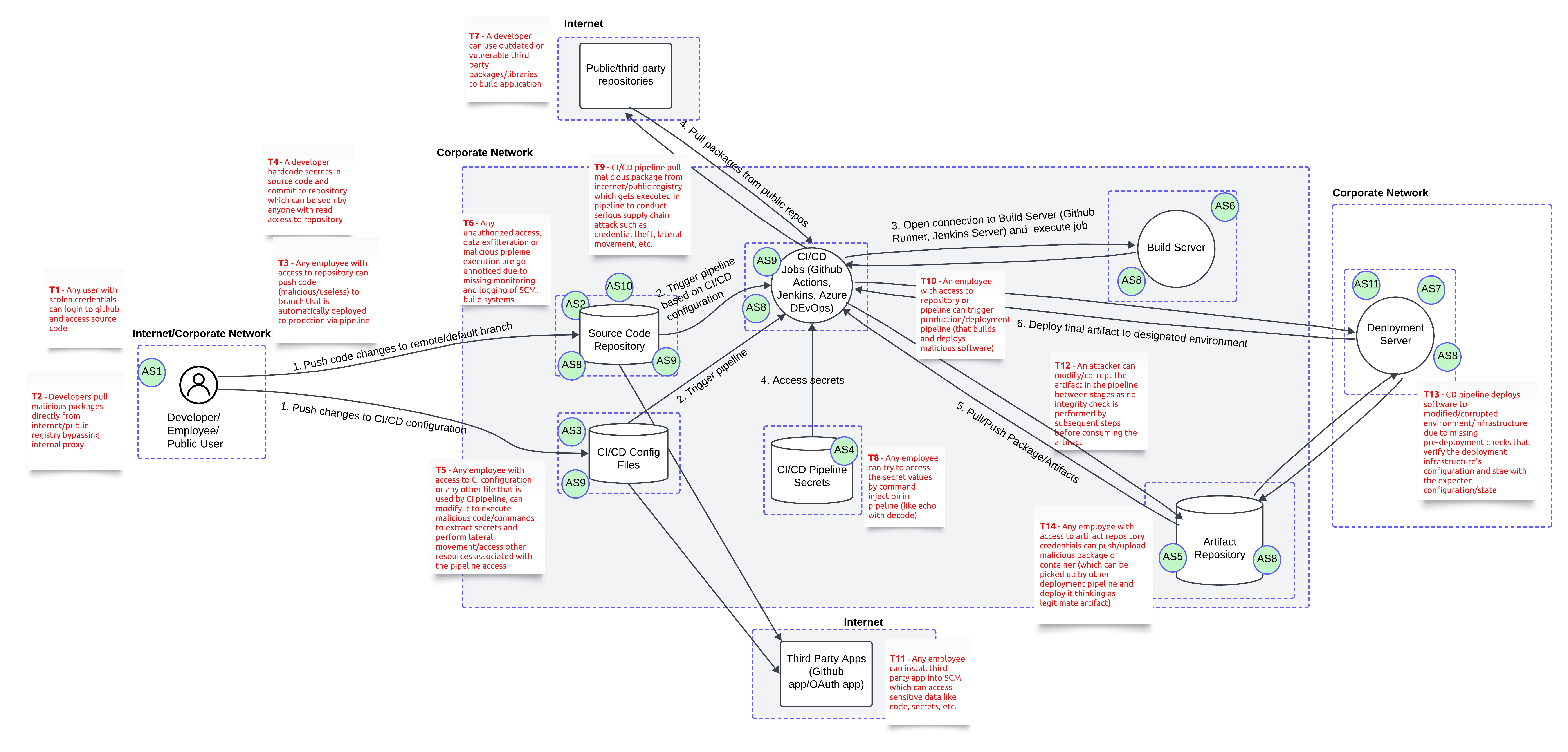}
    \end{adjustbox}
  }
  \caption{CI/CD Pipeline Data Flow Diagram}
  \label{fig:fig3}
\end{figure}

The DFD, shown in Figure~\ref{fig:fig3}, models the interactions and data flows between core CI/CD components, including developers, source code management (SCM) systems, CI runners, artifact repositories, and deployment environments. Key trust boundaries—such as external input, third-party tools, and internal services—are clearly delineated to support STRIDE-based threat analysis. Each actor and asset in the pipeline is assigned a unique identifier, which is referenced throughout the subsequent traceability matrix.

\subsection{Asset Inventory}
Based on the DFD, we identified critical assets that represent potential targets for adversaries. Each asset is assigned a unique identifier (e.g., \texttt{AS\#}) to facilitate cross-referencing in the threat traceability and control mapping tables that follow.

Table~\ref{tab:asset-inventory} lists the critical assets identified from the CI/CD pipeline Data Flow Diagram (DFD). 

\begin{table}[ht]
\centering
\caption{CI/CD Pipeline Asset Inventory}
\label{tab:asset-inventory}
\renewcommand{\arraystretch}{1.3}
\begin{tabular}{|p{2cm}|p{11cm}|}
\hline
\textbf{Asset ID} & \textbf{Asset Details} \\
\hline
AS1 & User Credentials (developers, admins, CI/CD systems) \\
\hline
AS2 & Source Code \\
\hline
AS3 & CI/CD Configuration Files and Scripts \\
\hline
AS4 & Secrets (e.g., environment variables, hardcoded credentials) \\
\hline
AS5 & Artifacts (e.g., binaries, images, packages, SBOMs) \\
\hline
AS6 & Build Machine and Data/Files Used During Build \\
\hline
AS7 & Deployment Infrastructure and Its Configuration \\
\hline
AS8 & Build and Audit Logs \\
\hline
AS9 & Access Control Policies \\
\hline
AS10 & SCM Metadata (commit history, user mappings, timestamps) \\
\hline
AS11 & Runtime Containers and Virtual Machines \\
\hline
\end{tabular}
\end{table}

These assets represent high-value targets that may be exposed to various threat agents across the pipeline lifecycle. In the next section, we analyze potential threats to these assets using STRIDE, and map them to specific security controls from SLSA and SSDF.

\subsection{Threat Agent Classification}
We classify potential threat agents based on their access level, intent, and position in the CI/CD pipeline. These actors range from fully external attackers to internal users with varying degrees of privilege as shown in Table~\ref{tab:threat-agents}. This classification supports targeted STRIDE threat modeling by helping identify plausible attack vectors for each CI/CD asset.

\begin{table}[ht]
\centering
\caption{Threat Agent Classification}
\label{tab:threat-agents}
\renewcommand{\arraystretch}{1.3}
\begin{tabular}{|p{2cm}|p{4.5cm}|p{7.5cm}|}
\hline
\textbf{Threat Agent ID} & \textbf{Threat Agent} & \textbf{Description} \\
\hline

TA1 & Public Malicious Actor & External attacker leveraging stolen credentials, exposed APIs, misconfigurations, or public interfaces \\
\hline

TA2 & Insider (Developer / Employee) & Authorized user who may act maliciously or accidentally compromise security (e.g., misconfigurations, secret leaks) \\
\hline

TA3 & Artifact Repository User & User with access to push or pull artifacts (e.g., container images, packages) from artifact repositories \\
\hline

TA4 & Build Runner User & User or automated process with access to CI/CD runners, build scripts, and temporary build artifacts \\
\hline

TA5 & Deployment Operator & User responsible for deploying builds into staging or production environments \\
\hline

TA6 & Source Control Admin (SCM Admin) & Admin-level user with elevated permissions in SCM platforms like GitHub or GitLab \\
\hline

TA7 & Artifact Repository Admin & Administrator with full control over the artifact repository (e.g., configuration, access control, cleanup) \\
\hline

\end{tabular}
\end{table}

\subsection{Threats-Security Controls Traceability Matrix}
The traceability matrix presented in Tables~\ref{tab:source-stride-checklist}, \ref{tab:build-stride-checklist}, \ref{tab:deployment-stride-checklist}, \ref{tab:monitoring-stride-checklist} maps CI/CD assets to STRIDE-based threats, associated threat agents, and applicable security controls. Controls are categorized according to both the Supply-chain Levels for Software Artifacts (SLSA) and NIST’s Secure Software Development Framework (SSDF), providing a layered and standards-based approach to software supply chain risk mitigation.

To further ground the threat scenarios in real-world attack patterns, each threat is also aligned with the most relevant category from the OWASP Top 10 — the industry-standard awareness document for common and impactful software vulnerabilities. This helps bridge abstract threat modeling with concrete software weaknesses, strengthening the rationale for selecting specific controls.

\subsubsection{Basis for Threat-to-Stage Mapping}
In addition to mapping threats to controls, each identified threat is associated with the most relevant CI/CD pipeline stage. This assignment follows a control-centric approach: the selected stage represents the most effective point of intervention — where preventive or detective measures can be applied to halt or reduce the threat’s impact.

While both the origin and potential propagation of the threat are considered, priority is given to where security controls (e.g., scanning, policy enforcement, gated approvals) can be practically integrated into the CI/CD flow. For example, a threat introduced in source code but realized post-build is assigned to the build stage, where it can be intercepted via verifiable builds or signature checks.

This mapping model supports actionable remediation by informing where controls are best implemented, in line with defense-in-depth principles. It complements the traceability matrix by aligning threats to their most effective mitigation stage within the pipeline lifecycle.

\subsubsection{Stage-Wise Threat–Controls Mapping}
Building on the control-centric threat-to-stage mapping described above, this section presents the detailed traceability matrix organized by CI/CD stages. Each subtable — Tables~\ref{tab:source-stride-checklist}, \ref{tab:build-stride-checklist}, \ref{tab:deployment-stride-checklist}, and \ref{tab:monitoring-stride-checklist} — focuses on one phase of the pipeline and lists:

\begin{itemize}
    \item The affected assets (AS\#)
    \item Relevant threat agents (TA\#)
    \item STRIDE threat categories (checklist format)
    \item Description of the threat scenario with OWASP Top 10 reference
    \item Applicable SLSA control(s)
    \item Applicable SSDF practice(s)
\end{itemize}
This structured format enables targeted implementation of security measures, aligned with real-world risks, established control frameworks, and the practical lifecycle phases of CI/CD. It facilitates measurable and prioritized hardening of the software supply chain.

\begin{enumerate}
    \item \textbf{Source Stage:} Threats originating at the source control level.
\begin{table}[htbp]
\centering
\caption{Source Stage Threat–Control Mapping with STRIDE Checklist}
\label{tab:source-stride-checklist}
\scriptsize
\begin{tabular}{|p{1.0cm}|p{1.0cm}|c|c|c|c|c|c|p{4.5cm}|p{5.2cm}|}
\hline
\textbf{Asset (AS\#)} & \textbf{Threat Agent (TA\#)} & \textbf{S} & \textbf{T} & \textbf{R} & \textbf{I} & \textbf{D} & \textbf{E} & \textbf{Threat Description (incl. OWASP)} & \textbf{Security Controls and SLSA/SSDF Mapping} \\
\hline

AS1 \newline
AS2 \newline
AS3 \newline
AS4 &
TA1 \newline
TA2 &
\ding{51} & \ding{55} & \ding{55} & \ding{51} & \ding{55} & \ding{51} &
\textbf{T1}: Stolen credentials used to access SCM. \newline OWASP: A01, A07 &
\textbullet\ MFA for all users \newline
\textbullet\ Rate limit login \newline
\textbullet\ RBAC (least privilege) \newline
\textbullet\ IDP-only login \newline
\textbf{SLSA:} None \newline
\textbf{SSDF:} PO.1.1, PO.2.1, PO.5.1, PO.5.2, PS.1.1 \\
\hline

AS2 \newline
AS3 \newline
AS5 \newline
AS6 \newline
AS7 &
TA2 \newline
TA6 &
\ding{51} & \ding{51} & \ding{51} & \ding{55} & \ding{55} & \ding{55} &
\textbf{T5}: CI/CD config tampering to inject secrets or logic. \newline OWASP: A08, A04 &
\textbullet\ Isolate CI config access \newline
\textbullet\ Block unreviewed runs (OPA) \newline
\textbullet\ Ephemeral runners \newline
\textbullet\ Branch protection \newline
\textbf{SLSA:} L3, L4 \newline
\textbf{SSDF:} PO.2.1, PO.2.3, PO.3.2, PO.4.2, PO.5.1, PS.1.1, PW.7.2, RV.3.2 \\
\hline

AS6 \newline
AS3 \newline
AS2 \newline
AS7 &
TA2 \newline 
TA6 &
\ding{55} & \ding{51} & \ding{55} & \ding{51} & \ding{55} & \ding{55} &
\textbf{T4}: Secrets hardcoded into code or artifacts. \newline OWASP: A02, A04 &
\textbullet\ Pre-commit secret scan \newline
\textbullet\ Repo scanning + rotation \newline
\textbullet\ Block deploy with secrets \newline
\textbf{SLSA:} None \newline
\textbf{SSDF:} PO.3.2, PO.4.2, PS.1.1, PW.4.1, PW.5.1, PW.7.2, RV.3.2, RV.3.3 \\
\hline

AS1 \newline
AS2 \newline
AS5 \newline
AS9 &
TA2 \newline
TA4 \newline
TA6 &
\ding{55} & \ding{55} & \ding{55} & \ding{51} & \ding{55} & \ding{51} &
\textbf{T11}: Developer installs malicious 3rd party GitHub/OAuth app. \newline OWASP: A05, A08 &
\textbullet\ Restrict app install to admins \newline
\textbullet\ Enforce GitHub org policy \newline
\textbullet\ Review app scopes \newline
\textbf{SLSA:} None \newline
\textbf{SSDF:} PO.1.1, PO.1.3, PO.2.1, PW.4.1, RV.1.1 \\
\hline

\end{tabular}
\end{table}
    \item \textbf{Build Stage:} Threats involving build systems, artifacts, and provenance.
    \begin{table}[htbp]
\centering
\caption{Build Stage Threat–Control Mapping with STRIDE Checklist}
\label{tab:build-stride-checklist}
\scriptsize
\begin{tabular}{|p{1.0cm}|p{1.0cm}|c|c|c|c|c|c|p{4.5cm}|p{5.2cm}|}
\hline
\textbf{Asset (AS\#)} & \textbf{Threat Agent (TA\#)} & \textbf{S} & \textbf{T} & \textbf{R} & \textbf{I} & \textbf{D} & \textbf{E} & \textbf{Threat Description (incl. OWASP)} & \textbf{Security Controls and SLSA/SSDF Mapping} \\
\hline

AS2 \newline
AS3 \newline
AS4 \newline
AS5 &
TA2 \newline
TA6 &
\ding{55} & \ding{51} & \ding{55} & \ding{51} & \ding{55} & \ding{51} &
\textbf{T3}: Developer pushes unreviewed code to auto-deploy branch. \newline OWASP: A08, A04 &
\textbullet\ Enforce PR review \newline
\textbullet\ Disable auto-merge \newline
\textbullet\ Manual approval required \newline
\textbf{SLSA:} L4 \newline
\textbf{SSDF:} PO.3.3, PW.5.1, PW.7.1, PW.7.2, PS.1.1 \\
\hline

AS5 \newline
AS3 &
TA2 \newline
TA3 \newline
TA7 &
\ding{55} & \ding{51} & \ding{51} & \ding{51} & \ding{51} & \ding{51} &
\textbf{T14}: Malicious artifact pushed or deleted from repository. \newline OWASP: A08, A05 &
\textbullet\ Restrict uploads to trusted pipelines \newline
\textbullet\ Attach and verify metadata \newline
\textbullet\ Artifact signing + delete restrictions \newline
\textbf{SLSA:} L2, L3 \newline
\textbf{SSDF:} PS.3.1, PO.3.2, PS.1.1, PS.3.1, PS.3.2, RV.1.3 \\
\hline

AS2 \newline
AS3 \newline
AS6 \newline
AS7 &
TA1 \newline
TA2 \newline
TA3 \newline
TA4 &
\ding{51} & \ding{51} & \ding{55} & \ding{51} & \ding{55} & \ding{51} &
\textbf{T9}: Malicious packages pulled from public registries (e.g., typosquatting). \newline OWASP: A06, A08 &
\textbullet\ Use internal proxies for dependencies \newline
\textbullet\ Enforce scoped package usage \newline
\textbullet\ Scan dependencies pre-commit \newline
\textbf{SLSA:} None \newline
\textbf{SSDF:} PO.3.1, PO.3.2, PO.4.1, PO.4.2, PO.5.1, PW.1.3, PW.1.4, PW.4.2, PW.4.4, PW.8.1, RV.1.1 \\
\hline

AS2 \newline
AS3 \newline
AS4 \newline
AS5 \newline
AS6 \newline
AS7 &
TA2 \newline
TA6 &
\ding{51} & \ding{51} & \ding{55} & \ding{51} & \ding{55} & \ding{51} &
\textbf{T2}: Developers bypass proxy and pull from untrusted public sources. \newline OWASP: A06, A08 &
\textbullet\ Block internet-bound pulls (egress firewall) \newline
\textbullet\ Force internal proxy in CI configs \newline
\textbullet\ Monitor for direct fetch attempts \newline
\textbf{SLSA:} None \newline
\textbf{SSDF:} PO.3.1, PW.1.3, PW.4.1, PS.3.1 \\
\hline

AS3 \newline
AS2 \newline
AS5 &
TA2 \newline
TA6 &
\ding{55} & \ding{51} & \ding{55} & \ding{51} & \ding{55} & \ding{51} &
\textbf{T8}: Secrets exfiltrated via echo/base64 commands in CI job. \newline OWASP: A01, A05 &
\textbullet\ Block unsafe patterns with OPA \newline
\textbullet\ Alert if secrets printed \newline
\textbullet\ Use short-lived/OIDC-based secrets \newline
\textbf{SLSA:} None \newline
\textbf{SSDF:} PO.3.1, PO.3.2, PO.4.2, PW.4.1, PW.6.1, PW.8.1, RV.1.2 \\
\hline

AS1 \newline
AS4 \newline
AS2 &
TA2 \newline
TA5 \newline
TA6 &
\ding{55} & \ding{51} & \ding{51} & \ding{55} & \ding{55} & \ding{51} &
\textbf{T7}: Developer includes outdated/vulnerable dependencies. \newline OWASP: A06 &
\textbullet\ PR scan with SCA tools (Mend/Snyk) \newline
\textbullet\ Maintain SBOM \newline
\textbullet\ Use vetted registries \newline
\textbf{SLSA:} L4 \newline
\textbf{SSDF:} PO.3.1, PO.4.1, PO.4.2, PO.5.1, PW.4.1, PW.5.1, RV.1.1, RV.2.1 \\
\hline

AS4 \newline
AS2 \newline
AS8 \newline
AS10 &
TA2 \newline
TA6 \newline
TA7 &
\ding{51} & \ding{51} & \ding{51} & \ding{55} & \ding{55} & \ding{55} &
\textbf{T12}: Attacker tampers artifact between pipeline stages. \newline OWASP: A08 &
\textbullet\ Enforce artifact signing (Sigstore) \newline
\textbullet\ SHA-256 validation + in-toto \newline
\textbullet\ Verify provenance before deploy \newline
\textbf{SLSA:} L3, L4 \newline
\textbf{SSDF:} PO.1.3, PO.3.2, PS.1.1, PS.2.1, PS.3.1, PS.3.2, PW.4.1, PW.4.4, PW.6.1 \\
\hline

\end{tabular}
\end{table}

    \item \textbf{Deployment Stage:} Threats related to environment misconfiguration, access abuse, or artifact trust during release.
    \begin{table}[htbp]
\centering
\caption{Deployment Stage Threat–Control Mapping with STRIDE Checklist}
\label{tab:deployment-stride-checklist}
\scriptsize
\begin{tabular}{|p{1.0cm}|p{1.0cm}|c|c|c|c|c|c|p{4.5cm}|p{5.2cm}|}
\hline
\textbf{Asset (AS\#)} & \textbf{Threat Agent (TA\#)} & \textbf{S} & \textbf{T} & \textbf{R} & \textbf{I} & \textbf{D} & \textbf{E} & \textbf{Threat Description (incl. OWASP)} & \textbf{Security Controls and SLSA/SSDF Mapping} \\
\hline

AS2 \newline
AS3 \newline
AS4 \newline
AS5 &
TA2 \newline
TA6 &
\ding{55} & \ding{51} & \ding{55} & \ding{55} & \ding{55} & \ding{51} &
\textbf{T10}: Unauthorized triggering of production deployment pipeline. \newline OWASP: A08, A05 &
\textbullet\ Enforce pipeline trigger approvals \newline
\textbullet\ Separate deploy privileges (SoD) \newline
\textbf{SLSA:} L4 \newline
\textbf{SSDF:} PO.2.1, PO.2.2, PO.5.1 \\
\hline

AS7 &
TA2 \newline
(DevOps Insider) &
\ding{55} & \ding{51} & \ding{55} & \ding{55} & \ding{55} & \ding{55} &
\textbf{T13}: Deployment to misconfigured or compromised infra (e.g., drifted Kubernetes cluster). \newline OWASP: A05 &
\textbullet\ Enable IaC drift detection \newline
\textbullet\ Block deploy on config mismatch \newline
\textbullet\ Use Checkov, Terraform drift tools \newline
\textbf{SLSA:} None \newline
\textbf{SSDF:} PO.1.2, PO.3.2, PO.4.2, PO.5.1, PW.6.1, PW.9.1, PW.9.2 \\
\hline

AS1 \newline
AS2 \newline
AS5 &
TA1 \newline
TA2 \newline
TA6 &
\ding{55} & \ding{51} & \ding{55} & \ding{51} & \ding{55} & \ding{51} &
\textbf{T14}: Lack of logging or monitoring in SCM/pipeline/repo leads to undetected tampering or data exfiltration. \newline OWASP: A10 &
\textbullet\ Enable audit logs in GitHub, CI, artifacts \newline
\textbullet\ Monitor runners with Falco \newline
\textbf{SLSA:} None \newline
\textbf{SSDF:} PO.3.2, PO.5.1, PO.5.2 \\
\hline

\end{tabular}
\end{table}

    \item \textbf{Monitoring Stage:} Threats tied to log tampering, undetected runtime attacks, and insufficient observability.
    \begin{table}[htbp]
\centering
\caption{Monitoring Stage Threat–Control Mapping with STRIDE Checklist}
\label{tab:monitoring-stride-checklist}
\scriptsize
\begin{tabular}{|p{1.0cm}|p{1.0cm}|c|c|c|c|c|c|p{4.5cm}|p{5.2cm}|}
\hline
\textbf{Asset (AS\#)} & \textbf{Threat Agent (TA\#)} & \textbf{S} & \textbf{T} & \textbf{R} & \textbf{I} & \textbf{D} & \textbf{E} & \textbf{Threat Description (incl. OWASP)} & \textbf{Security Controls and SLSA/SSDF Mapping} \\
\hline

AS1 \newline
AS2 \newline
AS5 &
TA1 \newline
TA2 \newline
TA6 &
\ding{55} & \ding{51} & \ding{55} & \ding{51} & \ding{55} & \ding{51} &
\textbf{T6}: Missing monitoring or logging allows pipeline abuse or tampering to go undetected. \newline 
OWASP: A10 – Insufficient Logging and Monitoring &
\textbullet\ Enable SCM, CI/CD, registry logs \newline
\textbullet\ Forward logs to SIEM \newline
\textbullet\ Monitor runners with Falco or similar \newline
\textbf{SLSA:} None \newline
\textbf{SSDF:} PO.3.2, PO.5.1, PO.5.2 \\
\hline

\end{tabular}
\end{table}

\end{enumerate}

\subsubsection{Summary of Threat–Control Traceability}

The stage-wise traceability matrix reinforces the importance of applying a layered defense strategy within CI/CD pipelines. By mapping threats to STRIDE categories and OWASP Top 10 weaknesses, and aligning them with corresponding SLSA and SSDF controls, organizations gain a multidimensional view of their software supply chain risks. This structured approach ensures that mitigation efforts are not only technically comprehensive but also contextually prioritized based on threat impact, exploitability, and control maturity. It highlights that while frameworks like SLSA provide verifiable integrity, and SSDF embeds secure development practices, threat modeling via STRIDE remains essential to ensure coverage of nuanced, real-world attack scenarios across all pipeline stages.

\section{Toolchain Integration for Pipeline Hardening}
Building upon the threat modeling and control identification methodology, this section presents a structured framework for integrating security controls within CI/CD pipelines. The framework aligns specific control objectives to relevant DevSecOps tooling and industry standards, enabling organizations to design defensible and standards-compliant software delivery workflows.

By aligning controls with established frameworks such as NIST SSDF and SLSA, this mapping provides a prescriptive reference for implementing supply chain security safeguards across pipeline stages.

Table~\ref{tab:toolchain-integration} illustrates representative mappings between threat scenarios, security control objectives, implementation tools, and applicable standards. These mappings are derived from the threat categories and control mechanisms outlined in previous stages and are intended to serve as practical guidance for integrating enforcement and observability tooling across CI/CD workflows.
\begin{table}[htbp]
\centering
\caption{Toolchain Integration for Threat-Driven Pipeline Hardening}
\label{tab:toolchain-integration}
\scriptsize
\renewcommand{\arraystretch}{1.3}
\begin{tabular}{|p{0.6cm}|p{4.2cm}|p{5.5cm}|p{2.5cm}|}
\hline
\textbf{T\#} & \textbf{Control Objective} & \textbf{Tool(s) and Framework(s)} & \textbf{CI/CD Stage(s)} \\
\hline

T1 & Prevent unauthorized access to SCM by enforcing identity verification and access control mechanisms & GitHub SSO, MFA, Branch protection, OIDC, Okta, Azure AD & Source \\
\hline

T2 & Restrict dependency resolution to approved internal repositories & Artifact proxy setup + policy enforcement in config files (\texttt{.npmrc}, \texttt{settings.xml}, etc.) & Build \\
\hline

T3 & Enforce pre-build validation gates and restrict automatic builds from unverified branches & PR approvals, branch whitelisting, OPA policies, signed commits (DCO/GPG) & Build \\
\hline

T4 & Prevent secret exposure in code repositories & Gitleaks, GitHub Push Protection, Vault, AWS Secrets Manager & Source \\
\hline

T5 & Enforce integrity and review of pipeline/IaC config files & GitHub Branch Protection, OPA, Checkov, KICS, CI linters & Source, Build \\
\hline

T6 & Enable centralized, tamper-evident logging and real-time monitoring & Falco, Osquery, GitHub/GitLab audit logs, ELK, Splunk, Wazuh, SIEM & Source, Build, Deployment, Runtime \\
\hline

T7 & Identify/remediate vulnerable OSS packages & Trivy, Mend, Snyk, Endor Labs, Dependabot, SBOM (Syft, CycloneDX) & Source, Build \\
\hline

T8 & Secure secret handling in pipelines and restrict exfiltration risks & Masked variables, job scoping, OPA command controls, audit logs & Build, Runtime \\
\hline

T9 & Detect and block vulnerable/malicious packages & SCA tools: Trivy, Mend, PR block policy & Build \\
\hline

T9 & Enforce allow-listing / proxy registry use & Internal proxy: Nexus, Verdaccio & Source, Build \\
\hline

T9 & Lock dependency versions and validate checksums & Lock files (e.g., \texttt{go.sum}, \texttt{package-lock.json}), checksum validation & Source, Build \\
\hline

T9 & Enable runtime protection and outbound network controls & Egress restrictions, sandboxing, Falco, Sysdig & Runtime \\
\hline

T10 & Restrict CI/CD job triggers to authorized personnel with policy-based approval for production & RBAC, deployment approvals, identity-based triggers, OPA & Deployment \\
\hline

T11 & Restrict and monitor 3rd-party app integrations & GitHub Allow Lists, OAuth policies, CI audit logs, SIEM alerts & SCM, Access Control \\
\hline

T12 & Ensure artifact integrity via cryptographic signing and policy enforcement & Cosign, in-toto, SLSA Provenance, GPG, OPA policies, access logs & Build, Deployment \\
\hline

T13 & Validate target infra config and detect drift before deploy & Driftctl, Terraform diff, OPA, tfsec, Checkov, kube-bench & Deployment \\
\hline

T14 & Enforce artifact signing and integrity verification & SLSA Provenance, Sigstore Cosign, in-toto & Build \\
\hline

T14 & Require strong authentication and RBAC for artifact repositories & IAM policies, scoped tokens, RBAC in Artifactory, ECR & Artifact Storage \\
\hline

T14 & Immutable artifact policies to prevent overwrites or deletion & Registry immutability/versioning settings & Artifact Storage \\
\hline

T14 & Audit logs and alerting on sensitive artifact operations & CloudTrail, ELK, Splunk & Artifact Storage, Runtime \\
\hline

\end{tabular}
\end{table}

This toolchain mapping serves as a practical reference for organizations aiming to harden their CI/CD pipelines against software supply chain threats. By integrating these tools, organizations can enforce preventive, detective, and corrective controls throughout their software delivery workflows.

Furthermore, the selection of tools aligns with NIST SSDF and SLSA, providing a standardized path toward higher levels of software supply chain integrity assurance. This methodology empowers security engineers to not only identify and assess risks through STRIDE-based threat modeling, but also to implement mitigations in a scalable, automated, and measurable manner—enabling DevSecOps practices grounded in threat-informed design.

\section{EVALUATION AND DISCUSSION}
\subsection{Framework Applicability and Security Coverage}
The proposed threat-driven security framework was evaluated based on its ability to identify, mitigate, and trace threats across the CI/CD pipeline. By combining STRIDE-based threat modeling with SLSA’s maturity levels and SSDF’s operational best practices, the framework ensures multilayered security coverage:

\begin{itemize}
    \item \textbf{STRIDE} enables structured identification of risks across pipeline stages and assets.
    \item \textbf{SLSA} introduces rigorous controls for build integrity, provenance, and artifact verification.
    \item \textbf{SSDF} incorporates people, process, and secure development practices often overlooked in technical control schemes.
\end{itemize}
The evaluation demonstrates that key threats—including source tampering, dependency confusion, credential leakage, and unauthorized pipeline execution—are effectively addressed through the traceability matrix and control mapping.
\subsection{Toolchain Readiness and Integration Feasibility}
Tools and platforms recommended in this framework (e.g., Sigstore, OPA, Trivy, Conftest, GitHub Policies) are widely available and compatible with modern CI/CD systems. Our analysis confirms:
\begin{itemize}
    \item \textbf{Incremental adoption is feasible}, allowing organizations to start with foundational practices (e.g., MFA, RBAC, SCA) and progress toward advanced capabilities (e.g., artifact provenance, policy-as-code).
    \item \textbf{Platform-neutral tooling} ensures integration with diverse CI/CD ecosystems like GitHub Actions, GitLab CI, Jenkins, and Kubernetes-native pipelines.
    \item \textbf{Policy-as-code} offers scalable enforcement, automatable validation, and auditability, aligning well with DevSecOps workflows.
\end{itemize}
\subsection{Discussion: Key Insights and Implications}
\textbf{1) Control-centric threat modeling improves prioritization.}  
By anchoring threats to the stage where controls are most effective, the framework helps teams focus on actionable mitigations and reduce blind spots (e.g., under-addressed runtime or configuration tampering threats).

\textbf{2) Standards alignment increases defensibility.}  
Mapping controls to SLSA and SSDF facilitates compliance reporting, internal audits, and alignment with federal and industry guidelines (e.g., NIST 800-218, Executive Order 14028).

\textbf{3) Multidimensional security is necessary.}  
No single framework alone (STRIDE, SLSA, or SSDF) provides end-to-end protection. The layered approach allows threat visibility, technical enforcement, and operational readiness to work in concert.

\subsection{Limitations and Practical Implications}
While the framework provides broad coverage, several practical considerations emerge:

\begin{itemize}
    \item \textbf{Tool complexity and cost.} Some advanced controls (e.g., ephemeral runners, full SBOM validation) may require additional engineering effort or licensed tools.
    \item \textbf{Threat landscape agility.} Supply chain attack vectors are evolving rapidly, requiring regular updates to threat models and control mappings.
    \item \textbf{Need for empirical testing.} Further testing in diverse organizational settings would strengthen the assurance of usability, scalability, and control efficacy.
\end{itemize}
These considerations highlight where ongoing refinement, organizational buy-in, and iterative deployment strategies are essential for effective CI/CD hardening.

\section{VIII. Conclusion and Future Work}

\subsection{Conclusion}

This paper presents a comprehensive, threat-driven framework to secure CI/CD pipelines against software supply chain attacks. By integrating STRIDE-based threat modeling with SLSA’s control maturity levels and SSDF’s operational best practices, the framework provides a multidimensional defense strategy tailored to the CI/CD lifecycle.

Key contributions include:

\begin{itemize}
    \item A structured methodology to identify and map threats across CI/CD stages using STRIDE and OWASP.
    \item A control-centric traceability matrix that aligns each threat with actionable mitigations based on SLSA and SSDF.
    \item Toolchain recommendations that operationalize controls through widely adopted, standards-aligned solutions.
    \item A methodology for contextualizing threat controls by CI/CD stage, maximizing the effectiveness of control placement.
\end{itemize}

This layered, standards-driven approach empowers organizations to move from ad hoc control deployments to defensible, auditable, and threat-informed pipeline hardening practices.

\subsection{Future Work}

Future work will focus on extending and validating this framework in real-world environments through:

\begin{itemize}
    \item \textbf{Case studies and pilot implementations} in diverse DevSecOps settings to assess performance, scalability, and usability.
    \item \textbf{Integration with threat intelligence platforms} to enable dynamic threat model updates based on real-time indicators.
    \item \textbf{Automation of threat-control mapping}, enabling continuous validation and policy enforcement in evolving CI/CD workflows.
    \item \textbf{Extension to runtime environments and software delivery ecosystems}, including package managers, container orchestrators, and cloud-native supply chains.
\end{itemize}

As software supply chain threats continue to evolve, proactive, threat-aware CI/CD security models like the one proposed here will be critical for ensuring trust, integrity, and resilience in modern software development.

\bibliographystyle{ieeetr}
\bibliography{references}

\end{document}